# Experimental confirmation of the low B isotope coefficient in MgB$_2$

P.Brotto[1], M.Tropeano[1], C.Ferdeghini,[1] P.Manfrinetti,[2] A.Palenzona,[2] E.Galleani d'Agliano,[1] and M.Putti[1]

[1]*CNR-INFM-LAMIA and Dipartimento di Fisica, Via Dodecaneso 33, 16146 Genova, Italy*

[2]*CNR-INFM-LAMIA and Dipartimento di Chimica and Chimica Industriale, Via Dodecaneso 31, 16146 Genova, Italy*

Recent investigations have shown that the first proposed explanations of the disagreement between experimental and theoretical value of isotope coefficient in MgB$_2$ need to be reconsidered. Considering that in samples with residual resistivity of few μΩcm critical temperature variations produced by disorder effects can be comparable with variations due to the isotopic effect, we adopt a procedure in evaluating the B isotope coefficient which take account of these effects, obtaining a value which is in agreement with previous results and then confirming that there's something still unclear in the physics of MgB$_2$.

Almost seven years after the observation of superconductivity in MgB$_2$ a full understanding of the physics of this material has not yet been achieved. First principle electronic structure calculations have shown the existence of two kinds of bands crossing the Fermi level (electron-like strongly anisotropic σ bands and hole- and electron-like 3D π bands). The existence of a strong electron-phonon coupling between Boron E$_{2g}$ modes and σ–band carriers has suggested that superconductivity in MgB$_2$ has an essentially phonon-mediated character. Due to the specific characteristics of the electrons involved in the coupling phenomenon, a peculiar multigap behaviour manifests, playing a crucial role in raising the Tc value up to 39 K[1,2]. Yet, some aspects of the physics of MgB$_2$ are not properly described in this framework and still stay under debate[3]. Among these a prominent position is occupied by the isotope effect, which was historically important in indicating the crucial role played by phonons in superconductivity and now, more than 50 years from its observation[4] it is still the key experiment to emphasize the conventional or unconventional nature of superconductivity in newly discovered superconductor materials.

The isotope coefficient, α, for a single element system with critical temperature T$_c$ and atomic mass *M*, is defined as:

$$\alpha = -\frac{d \ln T_c}{d \ln M} \qquad (1)$$

while for a multielement system the total isotope coefficient is just the sum over the individual atoms with mass $M_i$

$$\alpha = \sum_i \alpha_i = \sum_i -\frac{\partial \ln T_c}{\partial \ln M_i} \qquad (2)$$

The history of isotopic effect in $MgB_2$ was reviewed in 2003 by Hinks and Jorgensen[5]. The B isotope effect was first measured by Budko et al.[6] who found $\alpha(B) = 0.25(3)$ giving the first indication that phonons related to motion of B atoms, were involved in the pairing interaction. Hinks et al.[7] measured the isotope effect for both B and Mg confirming a large B isotope effect of 0.30(1), and a small effect for Mg, $\alpha(Mg)=0.02(1)$. The total measured isotope coefficient of 0.32 came out much less than the BCS value of 0.5, the expected value in the case of moderate coupling limit for a conventional superconductor.

This strong reduction of $\alpha$ could be in principle ascribed to one (or more) of the following features[1,2,3,5,8] : the Coulomb repulsion between paired electrons, the two-band character of $MgB_2$ superconductivity, and a large anharmonic character of the phonon spectrum (in particular of the $E_{2g}$ mode). Keeping into account the Coulomb repulsion in a simple one-band McMillan equation one reaches the conclusion[5] that unreasonably large values for the electron-phonon coupling constant $\lambda$ and for the Coulomb pseudopotential $\mu^*$ would be required to account for both $\alpha \approx 0.30$ and $T_c = 39$ K. A full two-band Eliashberg approach[1], which included an *ab initio* calculation of $\lambda$, and which used the "reasonable" value of $\mu^* = 0.12$ lead to $\alpha \approx 0.45$ and $T_c \approx 55$ K. Finally anharmonicity was proposed[1,2,5] as a possible explanation; it has the effect of increasing the relevant phonon frequency $\omega_{ph}$ of the $E_{2g}$ mode, with the effect of reducing $\alpha$, and of decreasing the coupling $\lambda$, with the effect of decreasing $T_c$. Detailed calculations performed using the frozen-phonon approach[1,8] showed that a 25% increase of the phonon frequency of $E_{2g}$ mode is expected as a consequence of the anharmonic effects. Such a value could explain the observed $\alpha$ and $T_c$ values and this seeming agreement between theory and experiment suggested that a global understanding of the electron-phonon coupling mechanism in $MgB_2$ had been achieved.

In a recent paper, however, Calandra et al.[3] reviewed the results of these calculations of the anharmonic phonon frequency shift, and showed that when all the leading order terms in anharmonic perturbation theory are included, the magnitude of anharmonic effects is marginal, invalidating the proposed explanation of the reduced isotope effect.

As suggested by the above discussion, the strong reduction of $\alpha$ is not yet well understood, and therefore, the isotope effect in $MgB_2$ is still an open question. On this ground, the aim of the present work is a very accurate new experimental investigation on the subject.

In fact, only now it has become manifest the significant effect of disorder on $T_c$. Irradiation experiments (for a review see ref. [9]) have shown that $T_c$ is reduced even by the smallest possible levels of disorder; a universal $T_c$ versus residual resistivity ($\rho_0$) relationship has been emphasized, which implies a reduction of $T_c$ comparable with the intrinsic variations due to isotopic effect for small $\rho_0$ variation less than 1 μΩcm. These results suggest that a sure evaluation of the isotopic effect should be done only in ideal defect-free samples. This is obviously impracticable, while a more realistic method is to introduce systematically small amount of defects in isotopically pure samples and to extrapolate $T_c$ for $\rho_0 \to 0$. This is the approach that we pursue in this communication.

We notice that, considering as a well established fact that the phonons involved in superconductivity in $MgB_2$ are primarily B phonons, we expect, in agreement with the previously quoted experimental result[7], a contribution from Mg to the total isotope coefficient much smaller than from B.

Therefore, in this work we focus our attention on the effect on the $T_c$ value of the B isotope substitution only.

For our study, $MgB_2$ samples were made with isotopic B ($^{11}$B and $^{10}$B enriched to 99.46% and 97.30%, respectively; Eagle-Picher) and natural Mg (Alfa Aesar 99.999% purity). The samples were produced by a single step technique[10], similar as in earlier work [6,11]. This technique provides dense (up to 2.4 g cm$^{-3}$, 90% of the theoretical density), clean and hard cylinder shaped samples with low residual resistivity ($\rho_0 \sim 0.5$ μΩcm) and high residual resistivity ratio (RRR~15). These values are indicative of the high purity of the phase and good connectivity between grains, which is crucial to have a reliable resistivity measurements.

In order to introduce disorder we followed different strategies. In the case of $Mg^{11}B_2$ the samples were irradiated with thermal neutrons at the Spallation neutron source SINQ at Paul Sherrer Institute with fluences varying in the range $10^{17} - 2 \cdot 10^{18}$ cm$^{-2}$. This method has proved to be able to introduce defects homogeneously in the samples and to vary systematically $T_c$ and resistivity[12,13]. Figure 1 shows the resistive transition versus temperature of this $^{11}$B sample series. It is seen that the critical temperature progressively decreases as far as the residual resistivity increases.

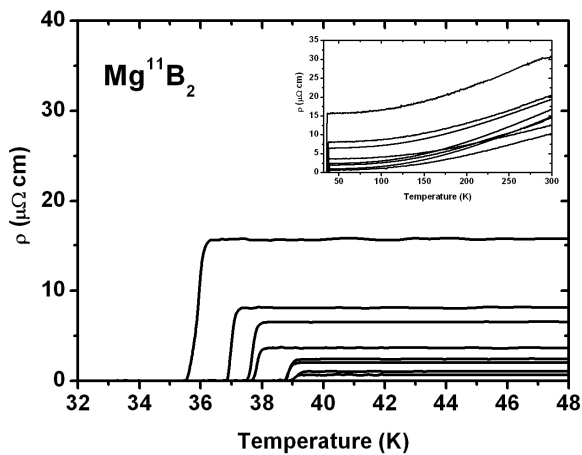

FIG. 1 Resistivity as a function of temperature for the $Mg^{11}B_2$ sample series. In the inset resistivity is shown up too room temperature.

In table I the main parameters of this sample series are summarized. It is interesting to note that the amplitude of the transition (ΔT~0.2-0.3 K) and the resistivity difference between room temperature and 41 K (Δρ ~ 9-15 μΩcm) remain nearly the same. This indicates that neutron irradiation produces a homogeneous defect structure and the connectivity between grains is not strongly affected[14].

TABLE I. Main parameters of the Mg$^{11}$B$_2$ sample series: residual resistivity $\rho_0 =\rho(41K)$, $\Delta\rho=\rho(300K)-\rho(41K)$, $\rho_{0,g} = \rho_0 \cdot \Delta\rho_g / \Delta\rho$ with $\Delta\rho_g$=7.5 μΩcm, T$_c$ evaluated at the 90% of the resistive transition, ΔT$_c$ evaluated between the 90% and the 10% of the resistive transition.

| Sample | ρ$_0$ (μΩcm) | Δρ (μΩcm) | ρ$_{0,g}$ (μΩcm) | T$_c$ (K) | ΔT$_c$ (K) |
|---|---|---|---|---|---|
| Virgin | 0.66 | 9.68 | 0.51 | 39.15 | 0.2 |
| Virgin | 1.02 | 13.7 | 0.56 | 39.20 | 0.2 |
| irradiated | 2.00 | 12.5 | 1.20 | 39.00 | 0.2 |
| irradiated | 2.40 | 14.3 | 1.26 | 39.05 | 0.3 |
| irradiated | 3.65 | 8.95 | 3.06 | 37.95 | 0.2 |
| irradiated | 6.50 | 12.9 | 3.78 | 37.80 | 0.2 |
| irradiated | 8.06 | 12.4 | 4.88 | 37.10 | 0.3 |
| irradiated | 15.66 | 15.0 | 7.85 | 36.10 | 0.3 |

The same technique does not apply to damage Mg$^{10}$B$_2$ samples since the huge cross section of the capture reaction, n+$^{10}$B, avoids the penetration of neutron over a thickness of about 40 μm from the surface. In this case a series of samples with different T$_c$ and resistivity values were obtained varying some preparation parameter, i.e. $^{10}$B particle size ( d < 22 μm and d < 50μm), and/or with subsequent annealing in dynamic vacuum and/or controlled atmosphere. Figure 2 shows the resistive transition of $^{10}$B sample series and in table II we report the main parameters; only samples in which a significant variation of T$_c$ and resistivity were observed are reported. Also this sample series presents sharp transition and good connection between grains except for the most annealed sample. In this case we have ΔT$_c$=0.5 K and Δρ=39.45 μΩcm, which implies a reduced connectivity of a factor five.

TABLE II. Main parameters of the Mg$^{10}$B$_2$ sample series : residual resistivity $\rho_0 = \rho(41K)$, $\Delta\rho = \rho(300K) - \rho(41K)$, $\rho_{0,g} = \rho_0 \cdot \Delta\rho_g / \Delta\rho$ with $\Delta\rho_g = 7.5$ $\mu\Omega$cm, T$_c$ evaluated at the 90% of the resistive transition, $\Delta T_c$ evaluated between the 90% and the 10% of the resistive transition.

| Sample | $\rho_0$ ($\mu\Omega$cm) | $\Delta\rho$ ($\mu\Omega$cm) | $\rho_{0,g}$ ($\mu\Omega$cm) | T$_c$ (K) | $\Delta T_c$ (K) |
|---|---|---|---|---|---|
| d< 22 µm | 0.58 | 7.94 | 0.55 | 40.20 | 0.2 |
| d< 22 µm | 0.58 | 7.91 | 0.55 | 40.25 | 0.2 |
| d< 50 µm | 2.34 | 8.48 | 2.07 | 39.40 | 0.2 |
| d< 50 µm | 3.00 | 8.19 | 2.75 | 39.20 | 0.2 |
| annealed | 5.54 | 14.9 | 2.79 | 39.20 | 0.2 |
| annealed | 22.00 | 39.5 | 4.18 | 38.65 | 0.5 |

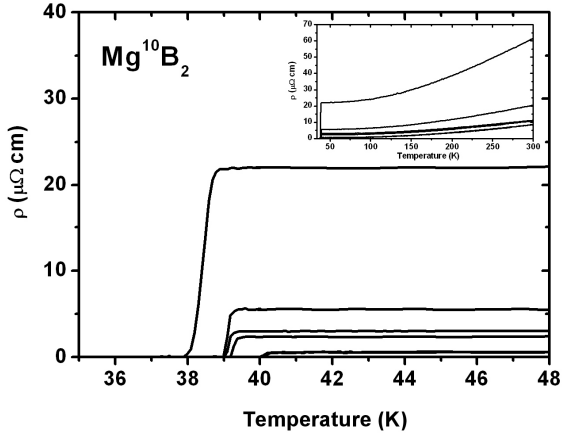

FIG. 2. Resistivity as a function of temperature for the Mg$^{10}$B$_2$ sample series. In the inset resistivity is shown up too room temperature.

To avoid problems related to poor connectivity, and extract the intragrain residual resistivity, $\rho_{0,g}$, the measured resistivity of the samples has been rescaled following the criterion proposed by Rowell[14]. In fact, neglecting grain boundary resistances, the measured resistivity $\rho$ is related to the intragrain resistivity $\rho_g$ by the following equation:

$\rho(T) = F[\rho_g(T) + \rho_{0,g}]$     $\Delta\rho = F\Delta\rho_g = \Delta\rho_g \rho_0 / \rho_{0,g}$

where $1/F$ is the fractional area of the sample that carries current, $\rho_g(T)$ is the temperature-dependent part and $\rho_{0,g}$ the residual part of intragrain resistivity, $\Delta\rho$ and $\Delta\rho_g$ are the changes in resistivity from 300 to 40 K. Within the Matthiessen rule, $\Delta\rho_g$ has a sample independent value, and the residual resistivity of grains, often indicated as $\rho_{0,g}$ can be estimated by the equation: $\rho_{0,g} = \rho_0 \cdot \Delta\rho_g / \Delta\rho$. According to a recent review[9] we have chosen $\Delta\rho_g$=7.5 µΩcm which is a typical value for connected thin films. Considering that $\Delta\rho_g$ plays here the role of a scaling parameter for the residual resistivity and we're interested in the extrapolation of $T_c$ value corresponding to $\rho_0$=0, the rather arbitrary choice of this parameter will not modify the main conclusion of this work.

In figure 3 $T_c$ vs $\rho_{0,g}$ is reported for the Mg$^{11}$B$_2$ and Mg$^{10}$B$_2$ series. The two sample series stay far from each other; each series shows a linear decreasing of $T_c$ with $\rho_{0,g}$ so that the data can be best fitted by a linear equation, y=ax+b. The best fitting parameters are the following: for the $^{11}$B series a= -0.44(2) K/µΩcm and b= 39.46(6) K; for the $^{10}$B series a= -0.44(2) K/µΩcm and b= 40.43(7) K. By definition the b parameter represents for each of the two series, $T_c$ in the limit of residual resistivity equal to zero. Thus from $T_c(\rho_0 = 0)$ =39.46(6) K for $^{11}$B and $T_c(\rho_0 = 0)$ =40.43(7) K for $^{10}$B we obtain from eq. (1) $\alpha$(B) = 0.264(3).

We point out that thanks to the procedure we used this evaluation is sample-independent and allows an intrinsic and definitive evaluation of the isotopic effect on B. The value we find is in substantial agreement with previous reports[6,7] and in particular it reproduces with higher precision the result in ref. [6].

Interestingly, in the two series of data $T_c$ decreases exactly with the same slope as a function of $\rho_{0,g}$ (the best fit parameter a is the same for the two series within the experimental indetermination). The two band model explains the suppression of $T_c$ with increasing disorder as an effect of interband scattering with impurities.[15] In particular, at low level of disorder, where other effects that can affect the density of states can be neglected[16], a linear relationship between $T_c$ and the interband scattering rate, is expected. Due to the multiband nature of MgB$_2$, this does not imply directly a linear relationship between $T_c$ and $\rho_0$; in fact residual resistivity is rather related to intraband scattering than interband ones, being the latter strongly suppressed[17]. However, $\rho_0$ is a good measure of disorder in the sample, and it is reasonable to assume that by increasing disorder interband scattering rate would increase proportionally with $\rho_0$. This can explain the linear suppression of $T_c$ with $\rho_0$[16], but it is still unclear why different sample series, which in principle present different nature of disorder, show the same $T_c$ vs $\rho_0$ slope. Such a universal behaviour, not expected in the case of multiband conduction, was recently pointed out also in ref. [18], indicating that this aspect needs more dedicated investigation.

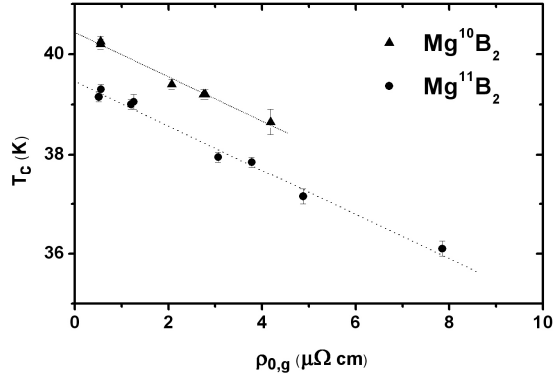

FIG. 3. Tc vs $\rho_{0,g}$ for the two series of samples. Data are reported in Table I and Table II. Error bars are reported taking ΔTc as the indetermination of experimental data. With this assumption we overestimate the real error but clearly show that the two series of data stay far away from each other.

Now we return to the discussion on isotopic effect. Our results, confirming definitely the previous experimental evaluations, put under discussion the present theoretical framework based on multi-band electron-phonon pairing.

One could thus be tempted to invoke other possible mechanisms, acting in addition to (or considerably modifying) the electron-phonon coupling pairing mechanism. This would be beyond the scope of the present work, even though it can be mentioned that, among all the proposed alternative mechanisms, there are both theoretical and experimental evidences indicating that non-adiabatic effects play a role[3,19] in $MgB_2$, and could therefore affect the superconducting pairing too.

A second aspect could be related to the different estimates given in the literature for both the effective electron-phonon coupling λ and the Coulomb pseudopotential $\mu^*$. As explained in ref [20] different choices of the couple of parameters λ and $\mu^*$ are possible, all of them reproducing the experimentally observed $T_c$. None of these choices seems in principle to be more plausible than another, and their possible relevance in relation to the isotope coefficient value has not yet been clarified. In particular, the first step in this direction could be an evaluation of the isotope effect in the framework of a more consistent treatment of the Coulomb repulsion[3].

Finally we would like to mention that recently a new mechanism has been proposed to play a role in $MgB_2$ superconductivity. The existence of low energy acoustic plasmons with sine like dispersion in $MgB_2$ has been theoretically predicted[21] suggesting that a plasmon mediated coupling could occur and this would explain the isotope coefficient reduction.

In conclusion we have confirmed the role of residual resistivity as a good "control-parameter" related to the amount of disorder in the samples, and we have surprisingly observed its independence on the nature of disorder introduced, which will be the object of further investigation. In this framework we have set an accurate procedure in order to obtain more certain experimental evaluations of intrinsic superconducting parameters of $MgB_2$ in which disorder effects are accounted for. This procedure has been adopted in the B isotope coefficient measurement. Our results give α(B) = 0.264(3) and, together with the smallness of the Mg isotope coefficient α(Mg) = 0.02, therefore confirm definitely a substantial reduction of the total coefficient α from the 0.5 BCS value. The anomalous isotope coefficient value emerges then as a still unresolved issue in the physics of MgB2, showing that there is something still unclear in the nature of superconductivity in this material.

This work is supported by MIUR under the projects PRIN2006021741. The authors thank S Massidda and G Profeta for fruitful discussions.

We also acknowledge the "Compagnia di S. Paolo" for financial support.

[1] H.J. Choi, D. Roundy, H. Sun, M.L. Cohen, S.G. Louie, Nature (London) **418**,758 (2002); H.J. Choi, D. Roundy, H. Sun, M.L. Cohen, S.G. Louie, Phys. Rev. B **66**, 020513(R) (2002)

[2] Y. Liu, I. I. Mazin, J. Kortus, Phys. Rev. Lett. **87**, 087005 (2001)

[3] M. Calandra, M. Lazzeri, F. Mauri, Physica C **456**, 38–44 (2007)

[4] E. Maxwell Phys. Rev. **78**, 477 (1950)

[5] D.G. Hinks, J.D. Jorgensen, Physica C **385**, 98–104 (2003)

[6] S.L. Bud'ko, G. Lapertot, C. Petrovic, C. E. Cunningham, N. Anderson, and P. C. Canfield, Phys. Rev. Lett. **86**, 1877 (2001)

[7] D.G. Hinks, H. Claus, J.D. Jorgensen, Nature **411**, 457 (2001)

[8] T. Yildirim, O. Gülseren, J.W. Lynn, C. M. Brown, T. J. Udovic, Q. Huang, N. Rogado, K. A. Regan, M. A. Hayward, J. S. Slusky, T. He, M. K. Haas, P. Khalifah, K. Inumaru, and R. J. Cava Phys. Rev. Lett **87**, 037001 (2001)

[9] M. Putti, R. Vaglio, J. Rowell, Supercond. Science and Tech. **21**, 043001 (2008)

[10] A. Palenzona, P. Manfrinetti and V. Braccini, INFM Patent No.T02001A001098

[11] Finnemore J. E. Ostenson, S. L. Bud'ko, G. Lapertot , and P. C. Canfield Phys. Rev. Lett . **86,** 2420 (2001)

[12] M. Putti, V. Braccini, C. Ferdeghini, F. Gatti, G. Grasso, P. Manfrinetti, D. Marré, A. Palenzona, I. Pallecchi and C. Tarantini, Appl. Phys. Lett. **86**, 112503 (2005)

[13] C. Tarantini, H. U. Aebersold, V. Braccini, G. Celentano, C. Ferdeghini, V. Ferrando, U. Gambardella, F. Gatti, E. Lehmann, P. Manfrinetti, D. Marré, A. Palenzona, I. Pallecchi, I. Sheikin, A. S. Siri, and M. Putti, Phys. Rev. B **73**, 134518 (2006)

[14] J. Rowell Supercond. Science and Tech. **16**, R17-R27 (2003)

[15] A. A. Golubov and I. I. Mazin Phys. Rev. B **55**, 15146 (1997)

[16] M. Putti, P. Brotto, M. Monni, E.Galleani D'Agliano, A. Sanna, S.Massidda, Europhys. Lett **77**, 57005 (2007)

[17] I.I. Mazin, O. K. Andersen, O. Jepsen, O. V. Dolgov, J. Kortus, A. A. Golubov, A. B. Kuz'menko, and D. van der Marel, Phys. Rev. Lett. **89**, 107002 (2001)

[18] M Eisterer Supercond. Science and Tech. **20**, R47-R73 (2007)

[19] M. Calandra and F. Mauri, Phys. Rev. B **71**, 064501 (2005)

[20] A. Floris, A. Sanna, M. Luders, G. Profeta, N.N. Lathiotakis, M.A.L. Marques ,C. Franchini , E.K.U. Gross , A. Continenza and S. Massidda , Phisica C **456**, 45-53 (2007)

[21] V.M. Silkin, A. Balassis, P.M. Echenique and E.V. Chulkov, arXiv: 0805.1558